\documentclass[runningheads]{llncs}
\usepackage[T1]{fontenc}
\usepackage{graphicx}
\usepackage{multirow}
\usepackage{booktabs}
\usepackage{amsmath}
\usepackage{amssymb}
\usepackage{hyperref}
\usepackage{comment}

\begin{document}
%
\title{Unraveling the Impact of Visual Complexity on Search as Learning}


%
%
\author{Wolfgang Gritz\inst{1}\orcidID{0000-0003-1668-3304} \and
Anett Hoppe\inst{1,2}\orcidID{0000-0002-1452-9509} \and
Ralph Ewerth\inst{1,2}\orcidID{0000-0003-0918-6297}}

\institute{TIB -- Leibniz Information Centre for Science and Technology, Welfengarten 1B, Hannover, 30167, Germany \and 
Leibniz University Hannover, L3S Research Center, Appelstraße 9A, Hannover, 30167, Germany
\email{wolfgang.gritz@tib.eu}
}

\authorrunning{W. Gritz et al.}

\maketitle              
\begin{abstract}
Information search has become essential for learning and knowledge acquisition, 
offering broad access to information and learning resources. 
The visual complexity of web pages is known to influence search behavior, 
with previous work suggesting that searchers make evaluative judgments within the first second on a page.
However, there is a significant gap in our understanding of how visual complexity impacts searches specifically conducted
with a learning intent.
This gap is particularly relevant for the development of optimized information retrieval (IR) systems that effectively support educational objectives.

To address this research need, we model visual complexity and aesthetics via a diverse 
set of features, investigating their relationship with search behavior during learning-oriented web sessions.
Our study utilizes a publicly available dataset from a lab study where participants learned about thunderstorm formation.
Our findings reveal that while content relevance is the most significant predictor for knowledge gain, sessions with less visually complex pages are associated with higher learning success.
This observation applies to features associated with the layout of web pages rather than to simpler features (e.g., number of images).
The reported results shed light on the impact of visual complexity on learning-oriented searches, informing the design of more effective IR systems for educational contexts. 
To foster 
reproducibility, we release our source code\footnote{\url{https://github.com/TIBHannover/sal_visual_complexity}}.
\end{abstract}

\section{Introduction}
The Internet's ubiquity and the abundance of online information have fundamentally altered knowledge acquisition.
Web search engines have become indispensable tools, providing the first access point to finding relevant information and resources that enable users to increase their knowledge~\cite{Ilahi2019Digital,Vrana2017Perspective}.
The process of effectively searching and judging 
the relevance of resources is an important part of the learning process.
Therefore, it is crucial to understand the factors that influence search behavior and learning outcomes 
to improve the effectiveness of learning in this environment~\cite{OBrien2022Effects,Pushparaja2021UserExperience}.

Web search sessions can encompass transactional, navigational, and informational purposes~\cite{Broder02taxonomy}.
The research field \textit{Search as Learning} (SAL) focuses on informational search, characterized by a user's intent to acquire knowledge from web pages.
Learning success is frequently measured as \textit{knowledge gain (KG)}, defined as the change in the learner's knowledge state after the search process~\cite{Hoppe2018,Machado2020}.
The relevance of search results is intricately tied to a user's current knowledge state and the potential \textit{KG} offered by the web pages retrieved.
Previous research has underscored the importance of understanding learning scopes and detecting learning needs.
For example, Yu et al.~\cite{Yu2021} introduced a novel set of resource-centric features, primarily focused on the linguistic complexity of web page content to enhance the predictive models. 

The impact of \textit{visual complexity} (VisCom) on learning outcomes, however, remains largely unexplored.
According to Lazard \& King~\cite{LazardKing2020Objective}, we refer to visual complexity as features measuring the amount and organization of visual information in a web page.
Liu et al.~\cite{Liu22roles} and Reinecke et al.~\cite{Reinecke13predicting} have highlighted the impact of users' immediate judgments about website aesthetics, particularly in relation VisCom.
However, little is known about how VisCom influences the learning process.
An intuitive assumption could be that web pages with high complexity might be perceived as less accessible and hinder learning, especially if the subject is unfamiliar with the topic.

In this paper, we investigate the relationship between \textit{VisCom} and \textit{KG} during web search.
We model \textit{VisCom} using a diverse set of features, which can be divided into four categories:
(1) \textit{HTML features}: statistical measures relying on the HTML code of a web page;
(2) \textit{visual features}: statistical measures relying on the rendered web page;
(3) \textit{layout features}: measures describing the web page layout utilizing the Vision-based Page Segmentation Algorithm (VIPS)~\cite{Akpinar13Vision,Cai03Extracting};
(4) \textit{aesthetic features}: measures related to the Gestalt laws adopted to web pages.
Using a publicly available dataset for exploratory search~\cite{lightning}, we address the following research questions:
(1) Can visual complexity serve as a predictor of a user's knowledge gain?
(2) Which specific visual complexity features are significant predictors of users' knowledge gain during web-based learning sessions?

The remainder of the paper is structured as follows:
Section~\ref{sec:rw} sums up the literature on \textit{VisCom} of web pages and the research related to SAL.
In Section~\ref{sec:method}, we detail the various methods employed to extract features from web pages and explain the process of aggregating these features for users.
Section~\ref{sec:eval} describes the conducted experiments.
Finally, we summarize our findings, and suggest future research directions in Section~\ref{sec:conclusions}.

\section{Related Work}\label{sec:rw}

\paragraph{Search as Learning:}\label{sec:rw:sal}
Search engines have typically prioritized results 
based on relevance, user satisfaction, and commercial aspects (e.g., advertisement placement). 
However, since search engines have become essential learning tools, they must also be optimized for knowledge acquisition~\cite{Broder02taxonomy}.
In the field of SAL, research focuses on understanding common user behavior and improving retrieval and ranking algorithms for learning~\cite{Ghafourian2023Ranking,Rokicki2022Learning}, e.g., for \textit{KG} prediction.
Vakkari~\cite{Vakkari16} surveyed features that can indicate a user's learning needs and their potential influence on knowledge acquisition 
during the search process.
Roy et al.~\cite{Roy20Exploring} observed that learning occurs at different times depending on prior knowledge.
This underscores the importance of modeling the knowledge state and its evolution during web browsing to understand learning processes~\cite{RULK,RULKKG,RULKNE,Zein2023Evolution}.
Research on investigating features that impact \textit{KG} can be divided into two main streams:
The first one focuses on the learning-related user behavior, e.g., based on analysis of input queries~\cite{Collins-Thompson16}, navigation logs~\cite{Eickhoff2014lessons}, and other behavioral features~\cite{Gadiraju2018}.

The second research direction explores what characteristics of web resources facilitate learning~\cite{Gritz2021,Otto2021,Yu2018,Yu2021}.
For example, Syed~and~Collins-Thompson~\cite{SyedC18} considered document-level features to improve learning outcome for short- and long-term vocabulary learning.
A psychological study found evidence that text-based web pages seem to have a more substantial influence on a user's \textit{KG}~\cite{Pardi20}.
In this context, Ghafourian et al.~\cite{Ghafourian2023Readability} and Gritz et al.~\cite{Gritz2021} examined readability metrics and textual complexity regarding their connection to user behavior and their influence on \textit{KG} prediction.
For their prediction task, Yu et al.~\cite{Yu2021} utilized a comprehensive set of features, which included HTML statistics.
Finally, Otto~et~al.~\cite{Otto2021} studied the effect on \textit{KG} prediction, when readability and linguistic features are complemented with multimedia features.
These studies contribute to a comprehensive understanding of how various factors can impact and forecast learning outcomes during web search.
However, the influence of \textit{VisCom} on  learning outcomes has been relatively little studied, presenting a significant research gap that our work aims to address.

\paragraph{Complexity of Web Pages:}\label{sec:rw:viscom}
The prediction of the \textit{VisCom} of web pages has long been a significant research focus, particularly in the commercial sector~\cite{Ethier06web,Marmat2023Influence}.
In recent research, the primary objective has shifted towards automating the \textit{VisCom} evaluation process, predominantly relying on feature-based methods.
These methods harness web page attributes to assess \textit{VisCom}, ranging from basic features like background color to more intricate aspects such as layout symmetry.
Ivory et al.~\cite{Ivory01Empirically} demonstrated that page-level features could predict ratings provided by experts, highlighting the varying importance of these features across different web page categories.
Wu et al.~\cite{Wu11Evaluating} extracted features from web pages and evaluated their impact on overall visual quality, with layout and text features emerging as the most influential.
Altaboli and Lin~\cite{AlTaboli11Investigating} compared features from previous studies with question-based ratings, finding a high correlation between feature-based and question-based ratings.
Harper et al.~\cite{Harper13Analysing} delved into the Document Object Model (DOM) of web pages to analyze \textit{VisCom}, revealing that structural aspects significantly influence user perceptions of complexity.
Reinecke et al.~\cite{Reinecke13predicting} emphasized the rapid formation of lasting judgments about a website's appeal within a split second of the initial view, with \textit{VisCom} emerging as a crucial factor due to its immediate perceptibility.
Their research underscored that \textit{VisCom} is more significantly influenced by broader layout features than tiny details.
Finally, Wan et al.~\cite{Wan21novel} analyzed how different layout aspects affect user perception and created an automatic prediction model for website aesthetics.



\section{Extraction of Visual Complexity Features}\label{sec:method}
In this section, we provide an overview of the collected features and how they were obtained.
A complete list can be found online\footnote{\url{https://github.com/TIBHannover/sal_visual_complexity/tree/main/appendix}}.

\subsection{Preprocessing}
\paragraph{DOM Tree Extraction:}
The Document Object Model (DOM) provides a hierarchical structure of the web page's HTML, with each node representing an element containing details like text, size, and style attributes.
We extract these nodes to analyze web page components systematically.

\paragraph{VIPS Algorithm Modification:}
We adapted the Vision-based Page Segmentation Algorithm (VIPS)~\cite{Cai03Extracting} to consider 
HTML5 elements following Akpinar and Yesilada~\cite{Akpinar13Vision}.
VIPS divides a web page into blocks based on its visual layout, providing a structured representation that enables deeper analysis of web complexity.

\subsection{HTML Features}
HTML features are derived from statistical analysis of the web page’s source code.
Using the \texttt{Beautiful Soup} library\footnote{\url{https://pypi.org/project/beautifulsoup4/4.12.2/}}, we extract elements such as text counts and styles.
Features include the number and distribution of specific HTML tags, which serve as proxies for the complexity and structure of the page.
The calculated features represent either the number, minimum, maximum, average, or standard deviation of occurrences of grouped HTML tags. 

\subsection{Visual Features}

The visual attributes of web pages are encapsulated within the visual features category, consisting of seven distinct features originally proposed by Wu et al.~\cite{Wu13Measuring}.
The selection of these features is motivated by their ability to convey essential visual characteristics of web pages.
\texttt{avg\_brightness}, \texttt{avg\_hue}, and \texttt{avg\_colorfulness} quantify the page’s color scheme by converting screenshots to HSV color space.
Brightness measures overall lightness, hue represents dominant colors, and colorfulness indicates vibrancy.
\texttt{png\_size} and \texttt{jpg\_size} reflect the file size of the rendered screenshots, where larger sizes suggest higher visual detail and complexity.
We normalize these sizes by dividing them by the total pixel count (width × height) to adjust for different page dimensions.
\texttt{page\_width}, \texttt{page\_height}, and \texttt{aspect\_ratio} measure the spatial layout of the page.
The aspect ratio (width/height) gives insight into how the content is distributed vertically and horizontally.
These features provide a concise summary of the visual complexity of web pages, helping to assess how visual elements may influence cognitive load during web search sessions.

\subsection{Layout Features}
%
Layout features capture the structural complexity of web pages by analyzing the arrangement of visual elements.
Using the VIPS algorithm, we derive 
a tree structure that 
reflects the hierarchical layout of the page.
\texttt{n\_vips\_non\_leaf\_nodes} and \texttt{n\_vips\_leaf\_nodes} represent the number of non-leaf and leaf nodes in the VIPS tree.
Non-leaf nodes correspond to content blocks that contain other elements, while leaf nodes represent 
elements without children.
A higher number of nodes indicates a more complex layout structure.
\texttt{text\_area\_to\_whole\_page} and \texttt{n\_texts\_to\_whole\_page} measure the proportion of the web page occupied by text.
These features assess how much emphasis the page places on written content compared to other visual elements, offering insight into the balance between text and imagery.
\texttt{n\_vips\_layers} quantifies the depth of the VIPS tree, which reflects how many times the layout has been subdivided into smaller blocks.
A deeper tree suggests a more intricate design with multiple layers of nested content.
These layout features provide valuable information about the structural organization of web pages, helping to assess how design complexity might impact user navigation and cognitive load during web search sessions.

\subsection{Aesthetics Features}
In the last category of features, which focuses on the aesthetics of web pages, there is a total of 14 features, all of which are related to features from Ling et al.~\cite{Ling2003Modelling}, adopted for web pages by Wan et al.~\cite{Wan21novel}.
The features consist of:
(1)~\textit{Balance} measures the difference in the distribution of object area in opposite page parts, including top vs. bottom and left vs. right balance.
(2)~\textit{Equilibrium} indicates how well the page layout is centered on the page itself, taking into account the center of mass along the x and y axes.
(3)~\textit{Symmetry} assesses the symmetry of the page in vertical, horizontal, and diagonal directions.
(4)~\textit{Sequence} analyzes the alignment of objects to facilitate eye movement, checking if the information is ordered according to reading patterns.
(5)~\textit{Cohesion} measures how similar the aspect ratio of objects and the pages are.
(6)~\textit{Unity} describes how well objects belong together and are perceived as a single entity.
(7)~\textit{Proportion} evaluates how close objects are to preferred proportional relationships, such as squares and rectangles.
(8)~\textit{Simplicity} reflects the directness and singleness of form, emphasizing the minimization of objects and alignment points.
(9)~\textit{Density} indicates how well the page is covered with objects compared to an optimal 50\% coverage.
(10)~\textit{Regularity} examines the uniformity of objects based on principles, including alignment regularity and spacing regularity.
(11)~\textit{Economy} measures how carefully and discreetly objects are used to convey a message as simply as possible.
(12)~\textit{Homogeneity} assesses how evenly objects are distributed among the four quadrants of the page.
(13)~\textit{Rhythm} measures the similarity of the areas and center distances of objects across the four quadrants of a page.
(14)~\textit{Order and complexity} represents the mean of the previous 13 features.

These features are calculated based on different categories of website objects, including leaf nodes, text leaf nodes, image leaf nodes, form leaf nodes, and other uncategorized leaf nodes.

\subsection{Additional Feature Sets}
To ensure a comprehensive analysis, we incorporate additional feature sets, recognizing that while VisCom is our focus, it is only one factor among many influencing web search and learning.
Previous studies have shown that textual complexity and readability affect search and learning behavior~\cite{Ghafourian2023Readability,Gritz2021}, as do the characteristics of user queries~\cite{Eickhoff2014lessons}.
Web content relevance is also a key factor. Below, we describe three additional feature sets that capture textual complexity, content relevance, and user query behavior.

\paragraph{Preprocessing of Web Page Texts:}
Our features which reflect the textual complexity and relevance of web pages are based on the textual content extracted from HTML files. 
However, the raw text from HTML files often contains labels for buttons, navigation bars or other artifacts.
We first use \texttt{Trafilatura}~\cite{trafilatura} 
to extract the main text of a web page.
However, on closer inspection, we still find artifacts in the extracted texts.
Therefore, we decided to filter further with \texttt{spaCy}\footnote{\url{https://spacy.io/}} and ensure that a text section contains at least one verb and one noun.

\paragraph{Textual Complexity (TexCom):}
To capture \textit{TexCom}, we leverage a total of 32 features, primarily based on text statistics.
We use the Python package \texttt{readability}\footnote{\url{https://pypi.org/project/readability/}} to collect features that cover readability grades (e.g., Flesch Reading Ease), sentence information (e.g., words per sentence), word usage (e.g., number of conjunctions), and sentence beginnings (e.g., conjunction).

\paragraph{Web Page Relevance (WebRel):}
Determining the relevance of a web page is inherently subjective to the learner's specific needs.
We define \textit{WebRel} in a topic-dependent manner, focusing on facts related to the topic that one might expect to find on a page.
Initially, we prompted ChatGPT 3.5 to provide us with 10 concise facts about the formation of lightning and thunderstorms.
Subsequently, we encoded the answers using a BERT-based model~\cite{reimers-2019-sentence-bert}.
Furthermore, for each web page, we encoded every extracted text with the same model and calculated the cosine similarity.
For each page, we determined the maximum value for each fact, aiming to represent whether a fact is mentioned on a page.

\paragraph{Query Features (Query):}
In addition to resource-specific features, we incorporate features based on user behavior.
Therefore, we define 11 features, such as the number of queries and the average query length.

\subsection{Feature Aggregation}
As learners probably visit a different number of web pages during the search session, we calculate the average of individual features across all visited content pages.
Content page means pages that are not search engine result pages or web pages which act as video containers (e.g., YouTube).
This applies to all features, except query features, as they are not specific to content pages.

\section{Evaluation}\label{sec:eval}
The evaluation is structured as follows:
First, we introduce the \textit{SaL-Lightning} dataset~\cite{lightning} in Section~\ref{sec:eval:dataset}, emphasizing its characteristics.
Second, we outline the experimental setup, formulating \textit{KG} prediction as a classification task in Section~\ref{sec:eval:setup}.
Next, in Section~\ref{sec:eval:results}, we highlight the impact of different feature sets, with a special emphasis on the \textit{VisCom} features.
We further explore the interplay between feature selection and classification outcomes, analyzing subsets of \textit{VisCom} features.
Additionally, we investigate the combination of \textit{VisCom} with \textit{WebRel} features.
Finally, to analyze the possible impact of individual features, we conduct a feature importance analysis.

\subsection{Dataset}\label{sec:eval:dataset}
In our evaluation, we use the publicly available \textit{SaL-Lightning} dataset on exploratory web search~\cite{lightning}.
Participants were tasked with learning as much as possible about the generation of lightning and thunder within a constrained time of 30 minutes.
Multiple-choice tests consisting of 10 identical questions were completed one week before and immediately after the search sessions.
The participants received compensation for their involvement. 

Our analysis focuses on web pages with relevant and substantial content for the research objectives.
Therefore, we excluded search engine result pages.
We further excluded pages that focus on video content, such as YouTube, because they act more as containers for videos, limiting \textit{VisCom} judgment.
The study included data from 114 participants; after filtering for content pages, we could use data from 112.
We use the test data to define \textit{KG} as:
$KG = \frac{|post|-|pre|}{N}$, 
where $|post|$ and $|pre|$ represent the number of correctly answered questions on post- and pre-test, respectively, and
$N=10$ is the total number of test items.
The study's pre-test scores averaged $0.52\pm 0.18$, indicating the initial level of knowledge.
Post-test scores showed an improvement, averaging $0.73\pm 0.16$.
Consequently, the average knowledge gain (KG) was calculated to be $0.22\pm 0.18$.

The participant pool was predominantly female, comprising $95$ females and $17$ males with an average age of $22.8\pm 2.8$ years.
Each participant engaged in the search session for an average duration of $25:37\pm 6:28$ minutes.
They visited a total of $16.1\pm 6.8$ web pages, including SERPs and video pages,
respectively $6.8\pm 3.8$, across $195$ web pages after filtering for content pages.

\subsection{Experimental Setup}\label{sec:eval:setup}
In our experiments, we address the task of automatically predicting a participant's \textit{KG} using a predictive model.
We formulate this task as a classification problem, which closely aligns with existing literature~\cite{Gritz2021,Otto2021,Yu2018}.
Therefore, we employed a stratified 10-fold cross-validation (90\% training, 10\% test data) to ensure reliable and consistent outcomes despite the limited amount of data available.
Additionally, we utilize feature selection and optimize hyperparameters.

\paragraph{Knowledge Gain Classes:}
Following the literature~\cite{Yu2018,Yu2021,Otto2021,Gritz2021}, we divide \textit{KG} into three classes: \textit{low}, \textit{moderate}, and \textit{high}.
Instead of predicting exact test scores, we aim to evaluate if features can predict more or less successful learning sessions.
We normalize \textit{KG} using z-scores ($\tilde{y} = \frac{y-\mu}{\sigma}$), where $y$ is the \textit{KG}, $\mu$ the mean, and $\sigma$ the standard deviation.
Scores below -0.5 are classified as \textit{low} (43 instances), above 0.5 as \textit{high} (28 instances), and the rest as \textit{moderate} (41 instances).

\paragraph{Classifiers:}
\begin{table}[t]
\setlength{\tabcolsep}{2.9pt}
\centering
\fontsize{9}{10}\selectfont
\caption{Macro F1-scores and micro accuracy for the \textit{KG} prediction task, averaged across all classifiers. Results are grouped by baselines (BL), full feature sets (Full Sets), including the features of Otto et al.~\cite{Otto2021} (MultiM), as well as our Query, \textit{TexCom}, \textit{WebRel}, and \textit{VisCom} features, and subsets of \textit{VisCom} with (\checkmark) and without feature selection. Significant improvements over the best baseline are highlighted in bold.}
\label{tab:prediction_results}
\begin{tabular}{llcccc}
\toprule
 & \textbf{} &  & \textbf{$\text{F}_1\text{-score}$} & \textbf{Accuracy} & \multicolumn{1}{l}{\textbf{}} \\
 & \textbf{Approach} & \textbf{F. Sel.} & \textbf{avg (macro)} & \textbf{avg (micro)} & \textbf{p-value} \\
\midrule
\multirow{3}{*}{\rotatebox[origin=c]{90}{\textbf{BL}}} & \textbf{Most Frequent} &  & 18.5 & 38.4 & \multicolumn{1}{l}{} \\
 & \textbf{Stratified} &  & 34.9 & 35.7 & \multicolumn{1}{l}{} \\
 & \textbf{Uniform} &  & 32.2 & 33.0 & \multicolumn{1}{l}{} \\
\midrule
 \multirow{5}{*}{\rotatebox[origin=c]{90}{\textbf{Full Sets}}} & \textbf{MultiM~\cite{Otto2021}} &  & 30.2 $\pm$ 4.0 & 32.4 $\pm$ 4.2 & 0.998 \\
 & \textbf{Query} &  & 35.5 $\pm$ 3.4 & 38.2 $\pm$ 3.5 & 0.574 \\
 & \textbf{TexCom} &  & 34.5 $\pm$ 4.1 & 37.4 $\pm$ 3.0 & 0.811 \\
 & \textbf{Page Relevance} &  & \textbf{44.5 $\pm$ 8.7} & \textbf{46.9 $\pm$ 6.8} & \textbf{0.005} \\
 & \textbf{VisCom} &  & 38.4 $\pm$ 4.4 & 40.4 $\pm$ 4.7 & 0.136 \\
\midrule
 & \textbf{VisCom$_{10}$ + Page Relevance} & \textbf{\checkmark} & \textbf{46.3 $\pm$ 1.7} & \textbf{47.7 $\pm$ 1.4} & \textbf{0.000} \\
 \midrule
\multirow{8}{*}{\rotatebox[origin=c]{90}{\textbf{Subsets}}} & \textbf{VisCom (Aesthetics)} &  & 38.8 $\pm$ 4.7 & 40.2 $\pm$ 3.7 & 0.106 \\
 & \textbf{VisCom (HTML)} &  & 36.1 $\pm$ 6.8 & 37.1 $\pm$ 6.9 & 0.700 \\
 & \textbf{VisCom (Layout)} &  & 29.8 $\pm$ 5.5 & 34.9 $\pm$ 4.4 & 0.969 \\
 & \textbf{VisCom (Visual)} &  & 35.0 $\pm$ 4.4 & 36.6 $\pm$ 3.5 & 0.906 \vspace{4pt} \\
 & \textbf{VisCom (Aesthetics)} & \textbf{\checkmark} & \textbf{40.2 $\pm$ 2.2} & \textbf{41.4 $\pm$ 2.0} & \textbf{0.002} \\
 & \textbf{VisCom (HTML)} & \checkmark & 38.5 $\pm$ 6.5 & 39.3 $\pm$ 6.4 & 0.353 \\
 & \textbf{VisCom (Layout)} & \checkmark & 29.0 $\pm$ 2.6 & 32.0 $\pm$ 3.0 & 1.000 \\
 & \textbf{VisCom (Visual)} & \checkmark & 32.1 $\pm$ 4.2 & 33.5 $\pm$ 4.3 & 0.992 \\
 \bottomrule
\end{tabular}
\end{table}

The prediction of knowledge gain depends on both feature quality and the choice of classification algorithms.
Rather than identifying the best classifier, our goal was to evaluate the impact of the features on the prediction performance.
To ensure comprehensive analysis, we used a variety of classifiers, including k-Nearest Neighbors~\cite{KNN}, Gaussian Naive Bayes~\cite{gaussiannb}, Decision Trees~\cite{DecisionTree}, Support Vector Machines~\cite{SVM}, Random Forests~\cite{RandomForest}, Gradient Boosting~\cite{gradientboosting}, Multilayer Perceptrons~\cite{MLP}, and AdaBoost~\cite{Adaboost} from the \texttt{scikit-learn} library~\cite{scikit-learn}.
This diverse set allows us to assess the robustness across different models.


\paragraph{Hyperparameter Optimization:}
The choice of hyperparameters can substantially impact the performance of classifiers.
To identify the best settings, we systematically explore various hyperparameter combinations using a grid search approach.
Due to our use of cross-validation for evaluation, the hyperparameters are determined within each iteration on the 90\% training data,
using an additional 3-fold cross-validation (67\% training and 33\% validation data).
Our optimization objective is the macro $\text{F}_1\text{-score}$, which averages the $\text{F}_1\text{-scores}$ across classes.
In this way, we aim to prevent overemphasizing class distribution to avoid optimizing parameters for the dominant class.

\paragraph{Feature Selection:}
When working with a dataset of 114 features for 112 subjects, feature selection can help to improve accuracy or find the most relevant variables.
Therefore, 
during each iteration of cross-validation, we calculate the Pearson correlation coefficient of each feature with all other features in the training split.
The $\gamma\in\{1,\,5,\,10,\,15,\,20\}$ features with the highest correlation to \textit{KG} are selected.
The value for $\gamma$ is picked in the hyperparameter optimization and can vary among the iterations of the cross-validation.

\paragraph{Metrics:}
We employ $\text{F}_1\text{-score}$, and accuracy as evaluation metrics.
These metrics are defined as 
$F_1 = \frac{2 \cdot TP}{2\cdot TP + FP + FN}$ and $Acc = \frac{TP+TN}{TP+FP+TN+FN}$,
where TP, FP, TN, and FN represent true and false positives and negatives, respectively.

\paragraph{Baselines:}
We use three baselines to set a lower bound for the classification task:
(1) \textit{Majority}, which always predicts the majority class,
(2) \textit{Stratified}, which predicts randomly based on class distribution, and
(3) \textit{Uniform}, which predicts randomly without considering distribution.
These baselines define the threshold for random guessing.
Additionally, we include the multimodal features (\textit{MultiM}) from Otto et al.~\cite{Otto2021}, inspired in part by the visual appearance of web pages.

\paragraph{Significance of Classifier Results:}
We evaluate each classifier individually, averaging the results to avoid reporting each metric separately.
To test if the classifiers outperform random guessing, we conduct a one-sided t-test comparing classifier accuracy to the best baseline (\textit{Most Frequent}, $38.4\%$) with $\alpha=0.05$.
To account for type I errors across multiple feature sets, we apply a Bonferroni correction, adjusting the significance level to $\alpha_{Bon}=\frac{\alpha}{n}$, where $n$ is the number of settings.

\subsection{Results}\label{sec:eval:results}
\paragraph{Different Feature Sets:}
First, we compare the results of the five different feature sets. The results are displayed in Table~\ref{tab:prediction_results}.
For statistical significance, we obtain $\alpha_{Bon}=\frac{0.05}{5}=0.01$.
When comparing the results of the different feature sets, it is evident that the classifier for \textit{VisCom} achieved better results than for \textit{MultiM}~\cite{Otto2021}, \textit{Query}, and \textit{TexCom}.
Particularly striking is the observation that the average accuracy for the three feature sets is even below the baseline \textit{Most Frequent}, which simply predicts the majority class (\textit{low}).
Additionally, it is noteworthy that the average accuracy for \textit{VisCom} is $40.4\%$, which is $2.0\%$ above the baseline, but this difference is not significant.

Interestingly, the \textit{WebRel} features surpass all other feature sets and baselines significantly, with an average accuracy of $46.9\%$.
This difference is statistically significant.
The potential applications of \textit{Large Language Models} such as \textit{ChatGPT} and their impact on traditional web search in educational contexts raise intriguing new research questions.
However, as the focus of this study is the influence of \textit{visual complexity}, it will be explored in future research.

The reasons that the results obtained for the VisCom features are not significant could be manifold.
Two possible considerations are that the feature set is too large for some classifiers.
Another could be that the visual complexity is an accompanying aspect, but not a primary factor.
To further investigate this, we next look at the impact of different subsets of \textit{VisCom} features, as well as the combination of \textit{VisCom} features with the significant additional \textit{WebRel} features.

\paragraph{VisCom Subset Analysis:}
In this experiment, we used the subsets of the \textit{VisCom} features as input for classification to investigate whether individual subsets have different impacts.
As we computed the results both with and without feature selection, there are eight settings, yielding $\alpha_{Bon}=\frac{0.05}{8}=0.00625$ for significance.

In the results without feature selection, it is noticeable that the outcomes for three out of four different subsets are weaker than the baseline \textit{Most Frequent}.
Only the features related to the aesthetics of the web page perform slightly better, with an accuracy of $40.2\%$, albeit not significant.
Moreover, the average accuracy is very similar to the results with all features, differing by only $0.2\%$.
This could be attributed to the fact that $\frac{70}{114}=61.4\%$ of all features pertain to aesthetics, and these features probably dominate the classification result.

Particularly, aesthetic features naturally exhibit high correlations among themselves since the calculations remain the same, although the features were calculated for different sets of objects on a web page.
The results might contain noise due to an abundance of similar features.
Therefore, we recompute the results with feature selection.
Upon examination of the outcomes, it is evident that the average accuracy for \textit{Aesthetics} and \textit{HTML} has increased, while the standard deviation has decreased, indicating more robust results.
However, for \textit{Layout} and \textit{Visual}, the average accuracy has decreased.
One potential explanation could be that \textit{Aesthetics} and \textit{HTML} consist of more features, possibly reducing noise by the feature selection.
Notably, significantly better results are achieved for \textit{Aesthetics} compared to the baseline, suggesting that the visual aesthetics of web pages could have a slight impact on learning outcomes.
The result could indicate that the design of the web page could have a greater influence on learning success than the actual sheer number of images or texts.

\paragraph{Combination of \textit{VisCom} and \textit{WebRel}:}
While the full set of \textit{VisCom} features did not show significantly improved results, it appears that at least a subset of features may have an impact.
The question arises whether combining \textit{VisCom} with other features could lead to improvements.
To explore this, we merge the best feature set (\textit{WebRel}) with the \textit{VisCom} features and conduct the experiment again.
It is important to note that the size of the feature sets varies substantially, with 10 and 70 features.
Therefore, we manipulate the feature selection so that in each iteration of the cross-validation, exactly $\gamma=10$ features are selected, ensuring that both feature sets consist of the same number of features.

Both the $\text{F}_1\text{-score}$ and accuracy have experienced a slight increase ($+0.8\%$).
Particularly notable is the significant decrease in standard deviation ($-7.0\%$ and $-5.4\%$).
All classifiers achieved an accuracy ranging from $45.5\%$ to $49.1\%$ for the combination of feature sets, compared to the previous range of $39.3\%$ to $57.1\%$ for \textit{WebRel} alone, indicating increased robustness.
This observation supports our hypothesis that \textit{visual complexity} could be a mediating factor influencing the prediction of \textit{KG} in addition to the relevance of web page content.

\paragraph{Feature Importance:}
\begin{figure}[t]
    \centering
    \includegraphics[width=0.95\linewidth]{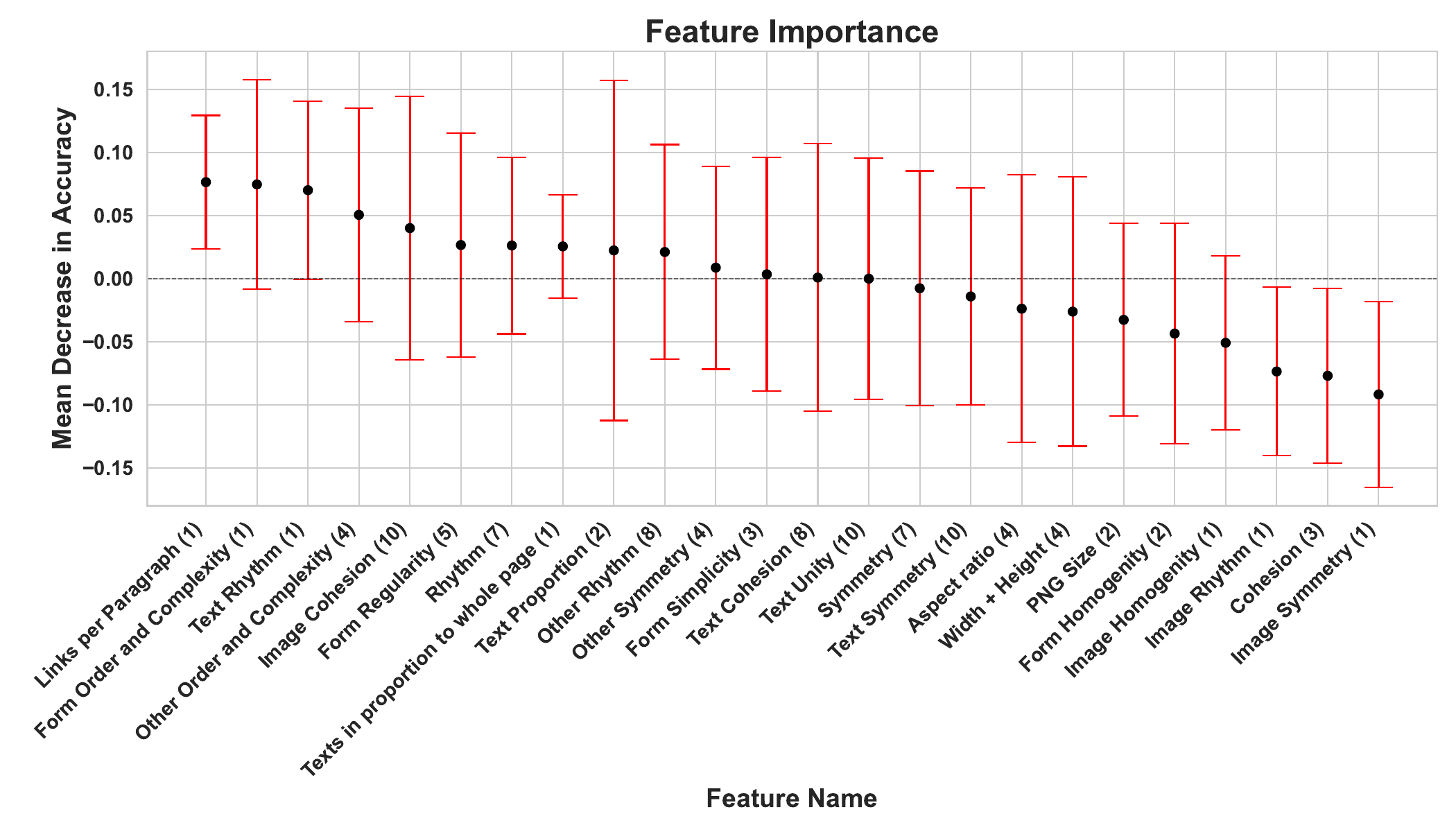}
    \caption{Permutation feature importance analysis for k-Nearest Neighbors, showing mean decrease in accuracy (black) and standard deviation (red) across cross-validation iterations. The selection frequency of each \textit{VisCom} feature is noted in brackets. Web page relevance features are omitted for clarity.}
    \label{fig:feature_importance}
\end{figure}
The considerably more robust results raise the question of which features had the greatest influence.
To analyze this, we calculated the \textit{permutation feature importance (PFI)} in each iteration of the cross-validation.
\textit{PFI} is a technique for determining the impact of individual features on the test result for various models.
A feature vector is randomly permuted, and the target metric (in our case, accuracy) is recalculated.
The \textit{PFI} for a feature is then defined as the difference between the original accuracy and the new accuracy with the manipulated feature vector:
$PFI(F_i) = \text{accuracy}_{\text{ori}} - \text{accuracy}_{\text{PFI}}$,
where $F_i$ is the considered feature, $\text{accuracy}_{\text{ori}}$ is the original accuracy, and $\text{accuracy}_{\text{PFI}}$ is the accuracy with the manipulated feature vector.
A higher value for \textit{PFI} could indicate that the feature has a positive impact on the outcome.
In each iteration of cross-validation, we perform the calculation for each feature 100 times to suppress random effects.
Note: Due to feature selection in each iteration, the selected \textit{VisCom} features may vary across cross-validation.

We investigate the influence of the combination of \textit{VisCom} and \textit{WebRel} from the previous experiment on the best classifier (k-Nearest Neighbors).
Subsequently, we calculate the average and standard deviation for each classifier across all iterations of cross-validation.
The sorted results for the \textit{VisCom} features are presented in Figure~\ref{fig:feature_importance} (\textit{WebRel} features are omitted for image clarity).
First, we observe that a total of 24 features were selected over the cross-validation iterations, with only three features selected in each iteration.
As the selection is based solely on Pearson correlation, potential causes could be fluctuations in correlation for different subsets of participants or strong correlations among the features.
Additionally, it is notable that only about half of the features exhibit a positive mean.
A negative value might imply that these features had a negative impact on the classification result, possibly due to the simplistic feature selection method.
Among the 12 more important features, 10 are \textit{Aesthetics} features, emphasizing their stronger influence as predictors,
emphasizing our observation that the layout of web pages has a more significant influence on learning success.


\paragraph{Limitations:}
Our findings are based on a dataset covering a single topic.
Thus, the results might not transfer to other populations or tasks.
However, as far as we know, this is the only dataset of an exploratory web search with a learning context in which the web pages were downloaded including all resources.
This allowed us to render web pages, which is crucial for most of our features.
To obtain more accurate and comprehensive results, the field requires larger studies covering longer periods~\cite{Bhattacharya2022Longitudinal}, different tasks, user groups, etc.

\section{Conclusions}\label{sec:conclusions}
In this paper, we investigated the effectiveness of \textit{visual complexity} features for \textit{knowledge gain} prediction using the 
\textit{SaL-Lightning}~\cite{lightning} dataset.
We defined and predicted knowledge gain as a classification task with three categories: low, moderate, and high, employing various classifiers, hyperparameter optimization, and feature selection.
Primarily, we analyzed the impact of different feature sets including \textit{textual complexity}, \textit{web page relevance}, \textit{query-based}, and \textit{multimodal features}~\cite{Otto2021} while paying special attention to \textit{visual complexity} features.

Our results showed that the complete set of \textit{visual complexity} features does not significantly outperform a random baseline.
However, testing the four subsets of visual complexity features revealed that the aesthetics subset yielded significant improvements, while the others (HTML, layout, and visual) had no notable effect.
This finding highlights the importance of website element arrangement in the learning process.
Moreover, combining \textit{visual complexity} features with \textit{web page relevance} -- which performed strongly across all feature sets -- enhanced the robustness of the classification results, suggesting a supporting role of \textit{visual complexity} in predicting \textit{knowledge gain}.
The feature importance analysis based on these combined feature sets further confirmed the pivotal role of \textit{aesthetics}.

These findings could be considered when designing teaching materials in order to support self-regulated learning and improve the effectiveness of knowledge acquisition.
Furthermore, existing search engines could explicitly include the visual complexity of web pages as an additional factor in the ranking of search results.
However, it must be mentioned that this study is limited to web search data on a single topic, thus future studies need to validate these findings for other topics.

{\fontsize{9}{11}\selectfont
\subsubsection*{Acknowledgments}
Part of this work was financially supported by the Leibniz Association, Germany (Leibniz Competition 2023, funding line "Collaborative Excellence", project VideoSRS [K441/2022]).
}

%
%
%
 \bibliographystyle{splncs04}
 \bibliography{bibliography}

\end{document}